# SURVIVAL ANALYSIS OF ORGANIZATIONAL NETWORKS – AN EXPLORATORY STUDY

Paula Lopes[1,2], Pedro Campos[1,2], Luís Meira-Machado[3]
[1]Faculdade de Economia, Universidade do Porto, [2]LIAAD INESC TEC
R. Dr. Roberto Frias, 4200-464 Porto, Portugal, Email: pcampos@fep.up.pt
[3]Department of Mathematics, University of Minho, Portugal, Email: lmachado@math.uminho.pt

Organizations interact with the environment and with other organizations, and these interactions constitute an important way of learning and evolution. To overcome the problems that they face during their existence, organizations must certainly adopt survival strategies, both individually and in group. The aim of this study is to evaluate the effect of a set of prognostic factors (organizational, size, collaborate strategies, etc.) in the survival of organizational networks. Statistical methods for time to event data were used to analyze the data. We have used the Kaplan-Meier product-limit method to compute and plot estimates of survival, while hypothesis tests were used to compare survival times across several groups. Regression models were used to study the effect of continuous predictors as well as to test multiple predictors at once. Since violations of the proportional hazards were found for several predictors, accelerated failure time models were used to study the effect of explanatory variables on network survival.

1. INTRODUCTION

Organizations are adaptive at different levels of analysis: individually, or as a group. In this work we dedicate a special attention to the network formation and regard networks as new forms of organizations. To understand the interactions between firms and the different mechanisms of adaptation, that include learning and evolution, we will use survival analysis, in particular to analyze the effects of a set of parameters (organizational density, size and age) in the founding and in the mortality of organizations.

An inter-organizational network is a set of firms that interact through inter-organizational relations (Eiriz, 2004). Johanson and Matsson (1987) describe the network as a system of relationships based on a division of work in the network. The notion of inter-organizational network is applied to a wide variety of relationships among organizations. The concept of inter-organizational networks can be applied to joint ventures, strategic alliances, industrial districts, consortia, social networks and others. Hakansson (1982, 1987), Hakansson *et al.* (1999) has studied the importance of relationships and learning in networks, giving important contributions to the study of inter-organizational networks. Some authors study certain properties of the network structure that are interesting to analyze. One of them is the *small-world* effect, which was brought to the field of economics by many researchers such as Watts and Strogatz (1997), Csermely (2006), Latora and Marchoiri (2003), etc. Watts and Strogatz (1998) have shown that the connection form of some biological, technological and social networks is neither completely regular nor completely random but stays somehow in between these two extreme cases. This type of networks is named *Small Worlds* in analogy to the concept observed in social systems by Milgram (1967). Small world networks are typically highly clustered like regular matrixes but have low path length, like random graphs. These properties are of great importance in economics. The work of Leskovec *et al*. (2005) confirm what was said about the low path length of the small-world networks. In their work, graphs densify over time (i.e., the number of nodes increase) and at the same time the average distance between nodes decreases. The diameter of small-world networks decrease as well along time. We distinguish three main forms that characterize the most common processes of cooperation: *linear*, *star* and *multipolar* networks. In a *linear* network (Figure 1, top), each of the nodes of the network is connected to two other nodes. All the flow that is transmitted between nodes in the network travels from one node to the next node in a linear manner. In this case, the activity $A_1$ is managed by firm $a_1$, activity $A_2$ is managed by $a_2$ and so on. No direction of the flows is identified in this graph, although the flow of resources can be unidirectional or bidirectional between nodes. Examples of this type of sequential networks correspond to the situations where firms collaborate with partners that are geographically close instead of searching for other networks that are already in action. The *star* network form (Figure 1, second row) corresponds to the type of network form in which each of the nodes of the network is connected to a central node. Usually, firms rationalize resources and optimize activities when they form this configuration. The sharing of new technologies from a particular organization and the common benefit of resources are the main advantages of the intervenient of these topologies. Here the flows can also be bidirectional because the sharing of resources is not centralized, but spread into the nodes of the network. Firms exchange, store or get resources via the central activity A. In many situations new firms are created to aggregate the activities of A: in this case, firm $a_5$ contains activity A. One example of networks that follow this form is the groups of suppliers. In this situation, firms are organized around a common client (usually a big dimension client). This form of organizational network represents an interesting opportunity for the cooperating firms to organize more efficiently their supplies. In such a situation the central activity is the goal of the network and constitutes the connection to the final client. The complementarities of the firms' competences are the key for the set of relationships that motivates this type of network, which is based on complementary of vertical relationships. Among the several advantages that firms can take from these groups of suppliers, we can emphasize cost reduction, access to new markets and risk reduction in the development of new products. The clustering of several organizations characterizes the *multipolar* network form (Figure 1, bottom). Typically, the relationships between organizations belonging to the same cluster are strong, but relationships between organizations that belong to different clusters are weak (although the relationships in this latter situation are stronger than with any organization that is outside the network). This kind of networks is typical from the automobile manufacturing sector although they can also be found in other industries, as for instance in textile and clothing industry.

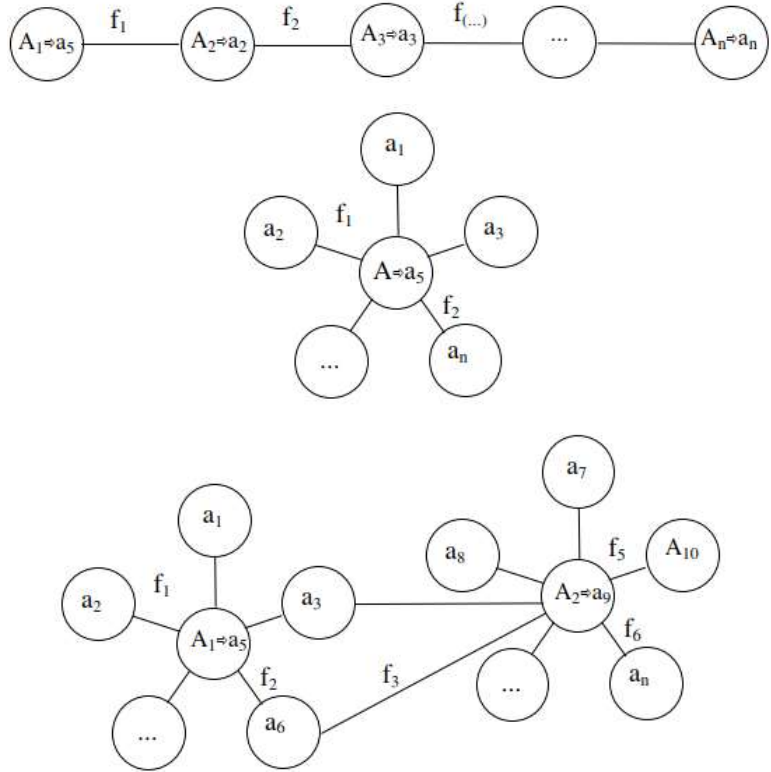

**Figure 1**: Representation of the simplified forms of networks: linear network (top), star network (middle) and multipolar network (bottom). Legend: $a_i$ denote the individual agents or firms in the network; $A_i$ denote the cooperating activities (integrated on firms) and $f_i$ denotes the flow of resources (no direction of the flow is represented).

Firms typically join forces following multiple cooperative strategies. According to Hitt *et al.* (2005), a network cooperative strategy covers situations where several firms agree to form multiple partnerships to achieve shared objectives. Doing better than competitors (through strategic execution or innovation) and merging and acquiring other companies are the two primary means by which firms develop a cooperative strategy. Following Hitt *et al.* (2005), cooperative strategies represent one of the major alternative firms use to grow. Besides the network form (*linear*, *star* or *multipolar*), the collaborative strategy adopted we consider the following main variables of analysis: profit, stock of knowledge, marginal cost, and some other variable concerning network statistics (number of networks, nodes, etc.), at the firm or network level. These variables were collected within each strategy. The presence of the binary variable status (taking the value 1 if the network is alive at the end of follow-up and 0 otherwise) makes the use of survival analysis the appropriate statistical methodology for analyzing the survival time of the network (Age). Survival analysis will be used to study the distribution of life time as well as to compare survival of two or more groups. Survival regression models will be used to study relationship between life time distribution and explanatory (covariate) variables associated with each network. In order to be able to analyze the importance of these variables on the survival of the networks, data of inter-organizational networks is needed. It is possible to know how many firms are born and die every year according to the number of firms that exist in the same year. Therefore, it is possible to compute the organizational density and measure its impact on actual organizational birth and death (contemporaneous density). Nevertheless, it is very difficult to analyze the impact of the founding density (the density at the time of founding) for a particular firm. For that, we should have real time-series for which we could capture information for every firm. That way we could follow firms during their lifetime. Information should contain several time periods for the same firm and it should also cover many observations, corresponding to different firms. The formation of networks also requires much information. Surveys can be implemented, in which we could collect data about the type of relationships and the reasons that force firms to link to networks. Yet to analyze the evolution of the network, several time points would be needed and therefore several interviews should be made. The shape of networks is also very dynamic and reality would be difficult to measure by the mean of surveys. It is also important to introduce different scenarios concerning the economic situation to analyze its impact in the evolution of firms and networks. For the reasons presented above, we have chosen to use simulation. Simulation is a simplification of the world, and a well-recognized way of understanding it. It provides tools to substitute for human capabilities, as the possibility of implementing different scenarios by constructing a virtual economic world. In this case we use microsimulation where the world events are driven by agent (firms) interactions. Emergent behavior of aggregate variables is then captured and the parameters can be reformulated in order to simulate different scenarios coming from different socio-economic perspectives. The rules that manage these interactions are described in Campos et. al. (2013).

## 2. SURVIVAL ANALYSIS OF ORGANIZATIONAL NETWORKS

This section analysis how a set of explanatory variables under study influence the survival of organizational networks. Survival analysis is the appropriate tool to model the lifespan of organizational networks. Conventional statistical techniques are not appropriate to model this type of duration data because the response variable (duration time) is not normally distributed and is partially observed because not all individuals (organizational networks) experience the event of interest (death) by the end of the observation period. This phenomenon, referred to as right censoring, must be accounted for in the analysis to allow for valid inferences.

In this study, simulated data from a set of 500 companies are studied from birth until death or end of study. The response variable is the age at which the company 'died'. At the end of study 122 companies where still active (i.e., 'alive') while 378 have 'died'. This means that we have 24.4% of right-censored observations in our data set. Besides the age or duration of life of the company we have several explanatory variables that will be used to model the duration of life of the network. Tables 1 and 2 present a short statistical summary of these variables. Except for the two nominal variables *form*, and *strategy* and for *status* (which is the state variable – with categories: death or alive - that determines the censored cases), all the rest are continuous or discrete attributes that can be treated as scale variables. Variable *form* indicates the shape of the network: it takes the value 1 when the shape is linear, the value 2 when it is a single star and the value 3 when it is a multipolar star (Figure 1).

Table 1: Distribution of the categorical variables

| Categorical variables | N (%) | Censored (%) |
|---|---|---|
| Final Status (*status*) | | |
| 0 – alive | 122 (24.4) | |
| 1 – dead | 378 (75.6) | |
| Form of the network (*form*) | | |
| 1- when the shape is linear | 352 (70.4) | 92 (26.1) |
| 2- when it is a single star | 111 (22.2) | 26 (23.4) |
| 3- when it is a multipolar star | 37 (7.4) | 4 (10.8) |
| Collaboration Strategy (*Strategy*) | | |
| A | 56 (11.2) | 18 (32.1) |
| B | 94 (18.8) | 18 (19.1) |
| C | 51 (10.2) | 22 (43.1) |
| D | 58 (11.6) | 8 (13.8) |
| E | 56 (11.2) | 12 (21.4) |
| F | 82 (16.4) | 20 (24.4) |
| G | 49 (9.8) | 10 (20.4) |
| H | 54 (10.8) | 14 (25.9) |

Table 2: Mean and standard deviation for the scale variables

| Variable | Mean (SD) |
|---|---|
| Age (age) | 4.914 (4.534) |
| Profit (profit) | 9.421 (4.457) |
| Marginal Cost (mcost) | 0.022 (0.023) |
| Number of existing networks at the time of its birth (netbirths) | 1.604 (0.675) |
| Number of existing networks at the time of its death (netdeaths) | 1.688 (0.709) |
| Number of existing nodes (firms) at the time of its birth (nodebirths) | 6.708 (4.091) |
| Number of existing nodes (firms) at the time of its death (nodedeaths) | 6.266 (3.891) |
| Stock of knowledge in Market X (stock1) | 8253.5 (31144.9) |
| Stock of knowledge in Market Y1 (stock2) | 17.163 (126.966) |
| Stock of knowledge in Market Y2 stock3 | 157.573 (1467.427) |

## 2.1. ESTIMATION OF SURVIVAL

Empirical estimation of the survival function can be obtained using the Kaplan-Meier estimator (Kaplan and Meier, 1958), also known as the product-limit estimator. The Kaplan-Meier estimator is the most widely used method to estimate the survival function. It is a nonparametric estimator that computes the probability of observing the event of interest at a certain point of time conditional to the survival up to that point. Figure 2 show a graphical representation of the Kaplan-Meier estimator of survival for the lifetime of the organizational network. The corresponding curve shows the estimated survival probabilities against time. It is a step function that starts at 1 at time 0 and that only decrease at event times. As shown in Figure 2, the survival function may not reach value 0 if the latest (largest) observations are censored. The Kaplan-Meier estimator can also be used to compute the median survival time, which is the time by when the probability of survival is 0.5. The median survival time can be appropriately estimated from the Kaplan-Meier curve as the x-axis (time) that crosses the horizontal line at the 50% survival probability on the y-axis. According to this method, the network median survival is equal to 4.

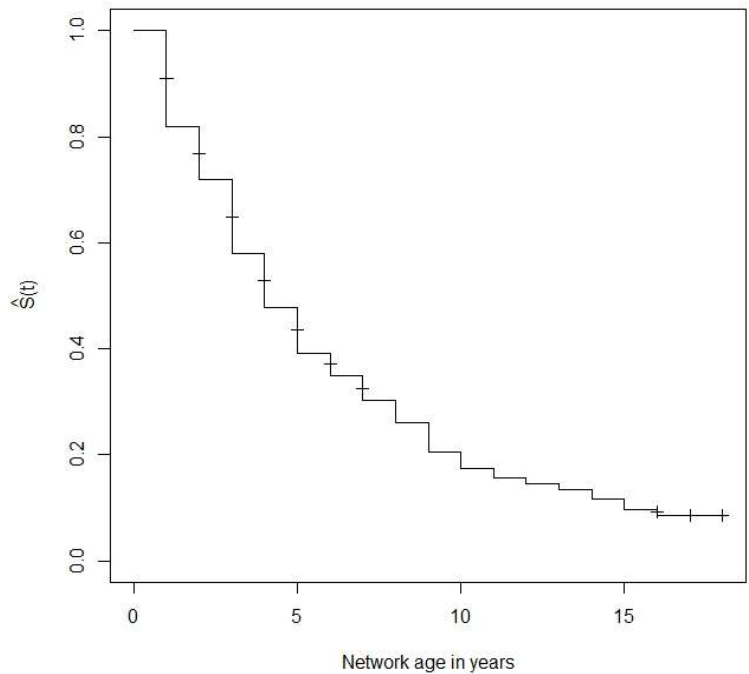

**Figure 2**: Kaplan-Meier estimator of survival.

Discrete covariates can be included in the Kaplan-Meier estimator by splitting the sample for each level of the covariate and applying the Kaplan-Meier method for each subsample. This is illustrated in Figures 3 and 4 for variables *form* and *Strategy*, respectively. It is obvious from the analysis of Figure 3 that those firms in a linear network (*form* =1) have poor survival than those in classified in a single star network (*form*=2) or in a multipolar star network (*form*=3). Table 3 reveals this issue too through the estimates of survival at 1, 3, 5 and 10 years since birth. Less differences can be appreciated when comparing the survival curves for *form* different than 1. In fact, it can be seen in Figure 3 but also in Table 2 that the two curves cross. The median survival for firms in a linear star network is equal to 3, where those values increase to 9 for those in a single star network and multipolar star network.

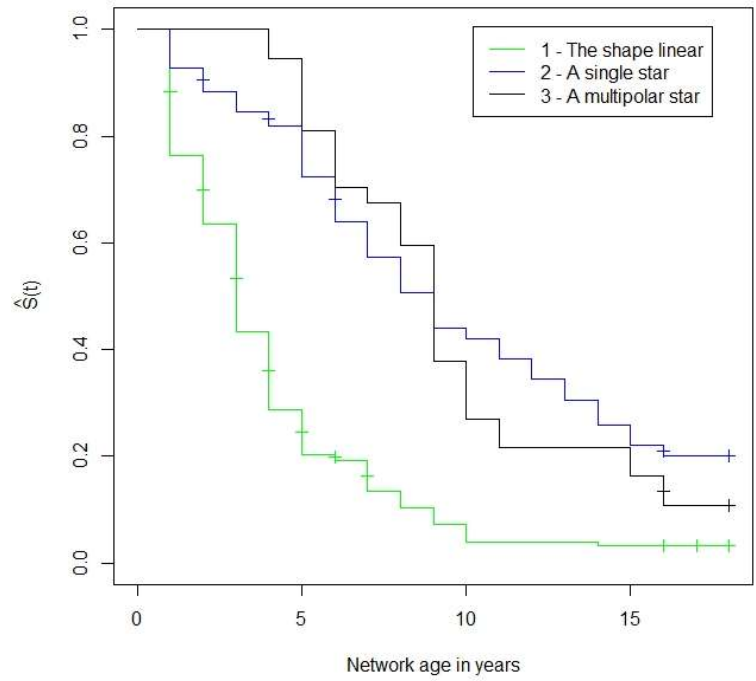

**Figure 3**. Kaplan-Meier survival curves for each level of the covariable *form*

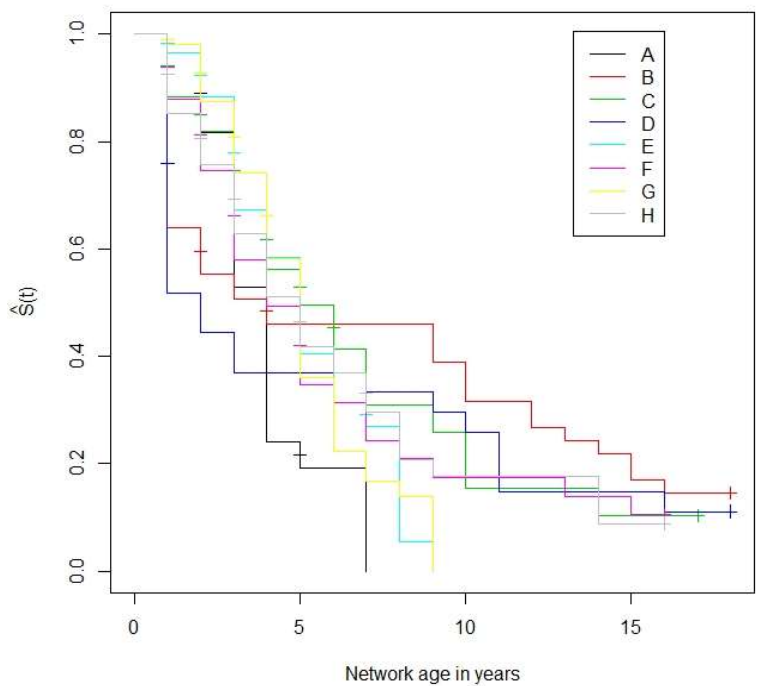

**Figure 4**: Kaplan-Meier survival curves for each level of the covariable *Strategy*.

**Table 3**: Survival estimates at 1,3, 5, 10 and 15 years for each level of variables form and Strategy.

|  | 1 | 3 | 5 | 10 | 15 |
|---|---|---|---|---|---|
| form | | | | | |
| 1-The shape linear | 0.764 | 0.434 | 0.204 | 0.039 | 0.031 |
| 2-A single star | 0.928 | 0.846 | 0.724 | 0.420 | 0.220 |
| 3-A multipolar star | 1.000 | 1.000 | 0.811 | 0.270 | 0.162 |
| Strategy | | | | | |
| A | 0.964 | 0.528 | 0.192 | - | - |
| B | 0.638 | 0.507 | 0.461 | 0.315 | 0.170 |
| C | 0.882 | 0.673 | 0.495 | 0.155 | 0.103 |
| D | 0.517 | 0.369 | 0.369 | 0.259 | 0.148 |
| E | 0.964 | 0.673 | 0.404 | - | - |
| F | 0.878 | 0.579 | 0.348 | 0.174 | 0.104 |
| G | 0.980 | 0.741 | 0.360 | - | - |
| H | 0.852 | 0.627 | 0.418 | 0.177 | 0.048 |

In Figure 4, we present the estimated probability of survival for each of the levels of the covariable strategy (*Strategy*) using the Kaplan-Meier estimator. In order to know if the survival of the organizational network can be influenced by the strategy, one can use a formal test to compare the survival distributions of the 8 samples. Under the assumption of proportional hazards, the log-rank test is optimal for testing the null hypothesis of equal survival distributions. However, as for covariable *form*, this assumption also seems to be violated for the covariable *Strategy* since several survival curves cross each other. Moreover, a test of the proportional-hazards assumption (Grambsch and Therneau, 1994) indicated that the assumption is indeed violated in this case ($\chi 2$ =108.837, P =1.6e-20). Under non-proportional hazards, log-rank is no longer the most powerful test (Li et al., 2015). The Peto & Peto (1972) modification of the Gehan-Wilcoxon method (Gehan, 1965; Harrington and Fleming, 1982) can be a good choice in this case since it gives more weight to events at early time points. Results for the two methods (log-rank test and the Gehan–Wilcoxon test) are shown in Table 4 revealing that they can lead to different conclusions.

**Table 4:** Tests for comparison of survival curves.

| | Log-rank | | Gehan-Wilcoxon | |
|---|---|---|---|---|
| Variable | $X^2(df)$ | p-value | $X^2(df)$ | p-value |
| form | 110 (2) | <2e-16 | 108(2) | <2e-16 |
| Strategy | 8.8 (7) | 0.3 | 14(7) | 0.05 |

Assuming the conclusion given by the Gehan-Wilcoxon test that at least one of these curves is different from the others, one naïve approach would be to perform pairwise comparisons. Results presented in Table 5 show the p-values from these comparisons with corrections for multiple levels using the one proposed by Benjamini and Hochberg (1995). The 8 groups lead to 28 pairwise comparisons with some p-values close to significance, in particular those involving group 1 and 4, corresponding to the strategies A and D. When confronted with a considerable number of curves one important question that tends to arise is if it is possible to group the curves in some manner. To this end we have applied to methodology proposed by Villanueva, Sestelo and Meira-Machado (2019) which allow us to ascertain whether these curves can be grouped or if all these curves are different from each other. Results from this methodology reveal the presence of three groups (i.e., strategies) with a similar survival pattern: Group 1 (A), Group 2 (B and D), Group 3 (C, E, F, G and H). The assignment of the curves to the three groups can be observed in Figure 4.

**Table 5**: P-values for the pairwise comparisons based on Peto & Peto test with corrections for multiple levels using the method proposed by Benjamini and Hochberg (1995).

```
    A     B     C     D     E     F     G
B 0.619 -     -     -     -     -     -
C 0.196 0.308 -     -     -     -     -
D 0.061 0.308 0.061 -     -     -     -
E 0.066 0.619 0.780 0.061 -     -     -
F 0.619 0.619 0.619 0.061 0.619 -     -
G 0.061 0.619 0.780 0.061 0.986 0.619 -
H 0.585 0.619 0.649 0.078 0.780 0.780 0.780
```

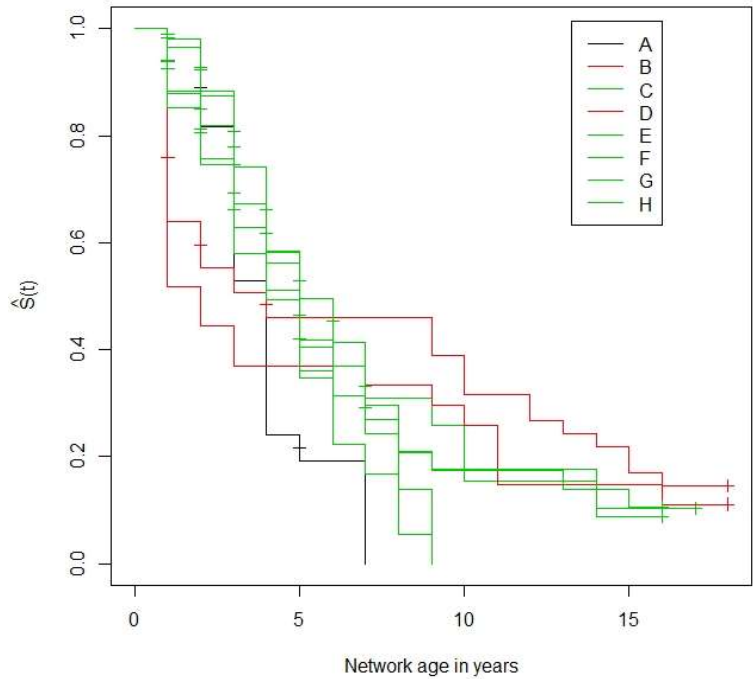

**Figure 5**: Estimated survival curves for each of the levels of the variable "Strategy". A specific color is assigned for each curve according to the group to which it belongs (in this case, three groups).

## 2.2. THE ACCELERATED FAILURE TIME (AFT) MODEL

Cox proportional hazards model (Cox, 1972) is the most used method for investigating the effect of explanatory variables on survival. A key assumption of the Cox regression model is that the hazard curves for the groups of observations should be proportional and should not cross. There are several graphical methods for identifying this violation, but the simplest is an examination of the Kaplan-Meier curves, such as those depicted in Figures 3 and 4. A more formal test was introduced by Grambsch and Therneau (1994). When there is clear evidence of nonproportional hazards, one should look for an alternative approach. Since the Accelerated Failure Time (AFT) models (Bagdonavicius, et al., 2002; Hougaard, 1999) do not exhibit proportional hazards, they can be considered as a good alternative to the Cox proportional hazards model. The AFT model describes a situation where the biological or mechanical life history of an event is accelerated (or decelerated). The hazard function of the AFT regression model can be written in the following form:

$$h(t|X) = h_0(t\exp(\beta^T X))\exp(\beta^T X)$$

where X is a vector of explanatory variables, $\beta$ is a vector of regression coefficients and $h_0(\cdot)$ is a baseline function of $t$, $X$ and $\beta$. Under the AFT model, the effect of the explanatory variables on the survival time is direct, accelerating or decelerating the time to death or failure.

The survival distribution for the AFT is given by

$$S(t|X) = S_0(t\exp(\beta^T X))$$

where $S_0(t)$ denotes the baseline survival function. The factor $e^\beta$ is called accelerated factor. This factor is the key measure of association obtained in the AFT model that can be used to evaluate the effect of predictor variables on survival time. Suppose we are considering a comparison of survival functions among two treatment groups. Then, the probability to survive time point $t$ in the treatment group is similar to the probability to survive time point $te^\beta$ in the control group.

The log-logistic and log-normal regression models are two of the most common examples of accelerated failure time models. The exponential and Weibull parametric regression models can be considered as AFT models too. In our analysis we considered several AFT models. Table 6 shows the models and the corresponding values for the Akaike's Information Criterion (AIC) which led us to choose the AFT model with log-logistic distribution as the best model.

**Table 6**: Akaike's Information Criterion values under different survival distributions for multivariable parametric survival regression models with covariates form, strategy, profit, netbirths, stock1 and stock2.

| Parametric Survival Model | AIC |
|---|---|
| Exponential | 2015.968 |
| Weibull | 1898.177 |
| Log-normal | 1847.173 |
| Rayleigh | 1928.973 |
| Log-logistic | 1842.017 |

Table 7: Univariable and multivariable Log-logistic survival regression models

| Covariable | Simple regression | | Multiple regression | |
|---|---|---|---|---|
| | $\hat{\beta}$ | P-value | $\hat{\beta}$ | P-value |
| Intercept | - | - | 1.310 | <2.0e-16 |
| Form | | | | |
| 1 – shape is linear | - | - | - | - |
| 2 – single star | 1.090 | <2.0e-16 | 0.997 | <2.0e-16 |
| 3 – multipolar star | 1.126 | <2.0e-16 | 1.170 | <2.0e-16 |
| Strategy | | | | |
| A | - | - | - | - |
| B | -0.009 | 0.963 | -0.511 | 1.7e-04 |
| C | 0.281 | 0.156 | 0.096 | 0.524 |
| D | -0.459 | 0.019 | -0.897 | 6.7e-10 |
| E | 0.226 | 0.205 | -0.190 | 0.166 |
| F | 0.110 | 0.515 | -0.156 | 0.241 |
| G | 0.229 | 0.208 | -0.310 | 0.026 |
| H | 0.164 | 0.382 | -0.361 | 0.010 |
| profit | 0.050 | 5.5e-06 | 0.031 | 5.1e-04 |
| mcost | -5.488 | 0.0039 | | |
| netbirths | -0.619 | <2.0e-16 | -0.211 | 5.1e-04 |
| netdeaths | -0.323 | 1.5e-06 | | |
| nodebirths | -0.016 | 0.190 | | |
| nodedeaths | -0.034 | 0.008 | | |
| stock1 | 2.88e-05 | 1.8e-06 | 1.4e-05 | 7.2e-04 |
| stock2 | 0.006 | 0.002 | 1.61e-03 | 0.176 |
| stock3 | 4.3e-04 | 0.081 | | |

Table 7 shows the results for the univariable and multivariable log-logistic regression models. Interpretation of the parameter estimates show us the impact of each prognostic factor on survival. A positive coefficient indicates higher survival times for the organizational network. For example, the coefficient for "form – single star" is 1.090, which indicates that the 'survival times' for organizational networks in this group are accelerated by a factor of 2.97 compared to those in group "form – shape is linear". Similarly, it can be seen that the survival times increases with an increase in the variables profit and stock1. On the other hand, the survival times decreases with an increase of netbirths.

## 3. CONCLUSIONS

A set of methods were used for analyzing the lifetime of organizational networks. The Kaplan-Meier method to estimate the probability of surviving beyond a certain time point, and hypothesis test to compare survival curves between different groups. In our study we were confronted with one exploratory categorical variable, strategy of the network, with a big number of levels. We have compared the corresponding survival curves through hypothesis tests and we conducted a study which established three groups of strategies with the same risk or survival probability.

Graphs for the estimated survival curves revealed that the survival curves for groups of observations cross each other revealing nonproportional hazards. More formal tests confirm this issue revealing that the well-known and widely used Cox proportional hazards model is not recommended to study the effects of the exploratory variables through regression studies. Accelerated failure time models based on different distribution were compared revealing that the log-logistic regression model to be the one with lower value for the Akaike's Information Criterion.